\documentclass[onefignum,onetabnum]{siamart190516}

\usepackage{lipsum}
\usepackage{amsfonts}
\usepackage{graphicx}
\usepackage{epstopdf}
\usepackage{algorithmic}
\usepackage[utf8]{inputenc}
\usepackage[english]{babel}
\usepackage[T1]{fontenc}
\usepackage{geometry}                % See geometry.pdf to learn the layout options. There are lots.
\usepackage[parfill]{parskip}    % Activate to begin paragraphs with an empty line rather than an indent
\usepackage{graphicx}
\usepackage{amsmath,amssymb,bbm, dsfont}
\usepackage[all,line]{xy}
\usepackage{epstopdf}
\usepackage{booktabs}
\usepackage{mathtools}
\usepackage{stackrel}
\usepackage{caption}
\usepackage{subcaption}
\usepackage{mathtools}
\usepackage{enumerate}

\ifpdf
  \DeclareGraphicsExtensions{.eps,.pdf,.png,.jpg}
\else
  \DeclareGraphicsExtensions{.eps}
\fi

\newsiamremark{remark}{Remark}
\newsiamremark{hypothesis}{Hypothesis}
\crefname{hypothesis}{Hypothesis}{Hypotheses}
\newsiamthm{claim}{Claim}

\headers{High-dimensional cointegration and Kuramoto systems}{J. Stærk-Østergaard, A. Rahbek, and S. Ditlevsen}

\title{High-dimensional cointegration and Kuramoto systems\thanks{Submitted to the editors \today.
\funding{Novo Nordisk Foundation NNF20OC0062958; 
Independent Research Fund Denmark | Natural Sciences 9040-00215B and Danish Council for Independent Research (DSF Grant 015-00028B)}}}

\author{Jacob Stærk-Østergaard\thanks{University of Copenhagen, Department of Mathematics.}
\and Anders Rahbek\thanks{University of Copenhagen, Department of Economics}
\and Susanne Ditlevsen\thanks{University of Copenhagen, Department of Mathematics (\email{susanne@math.ku.dk}).}
}

\usepackage{amsopn}
\DeclareMathOperator{\diag}{diag}

\newcommand{\CC}{{\mathbb C}}
\newcommand{\RR}{\ensuremath{\mathbb{R}}}
\newcommand{\eps}{\varepsilon}
\newcommand{\for}{\text{ for }}
\newcommand{\subspace}{\text{sp}}
\newcommand{\miss}[1]{{\bf\color{KUred} #1}}
\newcommand{\rank}[1]{ \text{rank}(#1)}
\DeclareMathOperator*{\tr}{Tr}
\DeclareMathOperator*{\abs}{abs}

\definecolor{KUred}{RGB}{144, 26, 30}
\definecolor{red1}{RGB}{220, 50, 50}
\definecolor{fxred}{RGB}{144, 26, 30}

\usepackage{fixme}
\fxsetup{
    status=draft,
    author=,
    layout=margin, % also try footnote or pdfnote, or inline
    theme=color
}
\definecolor{fxnote}{RGB}{144, 26, 30}

\ifpdf
\hypersetup{
  pdftitle={High-dimensional cointegration and Kuramoto systems},
  pdfauthor={J. Stærk-Østergaard, A. Rahbek, and S. Ditlevsen}
}
\fi

\begin{document}

\maketitle

% REQUIRED
\begin{abstract}
  This paper presents a novel estimator for a non-standard restriction to both symmetry and low rank in the context of high dimensional cointegrated processes. Furthermore, we discuss rank estimation for high dimensional cointegrated processes by restricted bootstrapping of the Gaussian innovations. We demonstrate that the classical \emph{rank test} for cointegrated systems is prone to underestimate the true rank and demonstrate this effect in a 100 dimensional system. We also discuss the implications of this underestimation for such high dimensional systems in general. Also, we define a linearized Kuramoto system and present a simulation study, where we infer the cointegration rank of the unrestricted $p\times p$ system and successively the underlying clustered network structure based on a graphical approach and a symmetrized low rank estimator of the couplings derived from a reparametrization of the likelihood under this unusual restriction.
\end{abstract}

% REQUIRED

% REQUIRED

%\cref{sec:conclusions}.

\section{Introduction}
High-dimensional data is ubiquitous in contemporary statistics with applications in many fields where laboratory equipment and techniques facilitate recording an ever increasing number of units simultaneously. In \cite{ostergaard2017} we explored a novel approach to infer network structure in a system of coupled oscillating processes by cointegration analysis. It was demonstrated how both uni-directional, bi-directional and all-to-all coupling was inferred. However, the studied systems only included four-dimensional processes which does not agree well with the interest in analysis tools for high-dimensional data series. 

In this paper we explore how cointegration analysis performs in high dimensions in terms of both rank and parameter estimation. We assume a high dimensional system of cointegrated processes with i.i.d. Gaussian innovations and a novel structure. To determine the rank of the matrix defining the couplings of the linear high dimensional system, we use restricted bootstrapping, as presented in \cite{cavaliere2012} for classical cointegrated systems, and adopt it to a high-dimensional setting. 

We furthermore define a linearized version of the classic Kuramoto model \cite{kuramoto1984} which implies a symmetric design of the couplings of the system. Estimation restricted to this non-standard condition requires development of a novel low-rank estimator of the system matrix under simultaneous restriction to both symmetry and low-rank. Here, we derive such an estimator by reparametrizing the likelihood function under symmetry restrictions. This estimator diverge from the standard maximum likelihood estimation using reduced rank regression. 

A simulation study highlights the performance of this estimator and discuss the implications of underestimating the true rank of the coupling matrix. Finally, using the estimated coupling matrix, we use a graphical method to infer the underlying cluster structure of the linearized Kuramoto system.

In the following, $I_p$ denotes the $p$-dimensional identity matrix and $M'$ denotes the transpose of a matrix $M$. For a real matrix $M\in\RR^{p\times r}$, $M_\perp\in\RR^{p\times (p-r)}$ denotes any matrix being an orthogonal complement, such that $M'_\perp M=0$, with $M$ and $M_\perp$ both of full rank. The matrix determinant operator is denoted $|\cdot|$. 
Finally, time is assumed discrete with positive index. Initial values are either assumed known or explicitly stated.

\section{Cointegration}\label{sec:cointegration}
Assume that $y_n=\left(y_{1n},...,y_{pn}\right)' \in R^{p}$ is a discrete time vector autoregressive process,
\begin{align}
	y_n = Ay_{n-1}+\mu+\eps_n,\label{eq:VAR1}
\end{align}
where $A\in\RR^{p\times p}$, $\eps_n$ is a $p$-dimensional Gaussian white noise process with covariance matrix $\Omega$ and $\mu\in\RR^p$ is a deterministic term. For simplicity, we assume $\Omega$ diagonal throughout the paper, such that all dependencies between variables are captured in the autoregressive matrix $A$. The \textit{characteristic polynomial} for (\ref{eq:VAR1}) is the determinant of $I_p-A\zeta$ for $\zeta\in\CC$. If the roots of the characteristic polynomial are all outside the unit circle, then the initial values of $y_n$ can be given a distribution such that $y_n$ is stationary, see \cite{johansen1996}.

If the characteristic polynomial of (\ref{eq:VAR1}) contains the root $\zeta=1$, then there exists no stationary solution of $y_n$, and we say that the process is \textit{integrated}. In particular, see \cite{johansen1996}, $\Pi=A-I_p$ will have reduced rank $r<p$ and can be written as $\Pi= \alpha\beta'$. The process $y_n$ is then integrated of order one, $I(1)$, with $r$ cointegrating relations $\beta'y_n$ under regularity conditions presented in Section \ref{sec:cointProcess}. 

In this paper we will only deal with $I(1)$ processes, so when we refer to $y_n$ as integrated, we implicitly mean that $y_n$ is integrated of order 1.

%%%%%%%%%%%%%%%%%%%%%%%%%%%%%%%%%%%%%%%%%%%%%%%%%%%%%%%%

\subsection{Cointegrated process}\label{sec:cointProcess}
Consider first the well-known results from cointegration analysis in \cite{johansen1996}. Rewrite (\ref{eq:VAR1}) with $\Pi=A-I_{p}$ in the \emph{vector-error-correction-model} form as
\begin{align}
	\Delta y_{n}=y_{n}-y_{n-1}=\Pi y_{n-1}+\mu +\varepsilon _{n}.\label{eq:VECMrep}
\end{align}
If $| I-A\zeta| =0$ implies $|\zeta|>1$ then $y_{n}$ has a stationary representation as an $I(0)$ process. In particular, $\Pi$ has full rank $p$ and all linear combinations of $y_n$ are stationary. If the $(p\times p)$-dimensional matrix $\Pi$ has reduced rank $r<p$ then $\Pi =\alpha\beta'$ with $\alpha,\beta,\enskip p\times r$ dimensional matrices of rank $r$. Moreover, the process $y_{t}$ is $I(1)$ with $r$ cointegrating stationary relations $\beta'y_t$ provided that the spectral radius $\rho(I_{r}+\beta'\alpha) <1$. This we refer to as the $I(1)$ conditions in the following. 

Note that if $r=0$ the process $y_n$ is $I(1)$ with a linear trend if $\mu\neq0$, but with no cointegration, while if $r=p$ (and $\rho(I_p+A) <1$) then $y_n\,$is $I(0)$, and $p$ stationary linear combinations exist. Under the reduced rank $r$, the system can be written as,
\begin{align*}
	\Delta y_{n}=\alpha\beta^{\prime }y_{n-1}+\mu +\varepsilon_n ,
\end{align*}
with $\beta$ containing the $r$ cointegration vectors and $\alpha$ the \emph{loadings} or \emph{adjustment coefficients}.

Note that the entries of $\alpha$ and $\beta$ are not uniquely identified, since we can use any non-singular transformation to obtain similar results. Rather, we identify the subspaces $\subspace(\alpha),\subspace(\beta)\in\RR^r$, that is, the subspaces spanned by the columns of $\alpha,\beta$. We can therefore choose the normalization 
\begin{align*}
	\beta^* = \beta(c'\beta)^{-1}, \quad\text{with } c = (I_r, 0_{p-r\times r})'
\end{align*}
of $\beta$ in order to identify parameters uniquely, given that the normalization is empirically valid. %A necessary condition for an $I(1)$ process is that  $|\alpha_\perp'\beta_\perp|\neq0$.

\subsection{Estimation of parameters}
Consider observations $(y_1,\dots,y_N)$ from the process (\ref{eq:VAR1}) and denote by $H_r$ the hypothesis $H_r: \rank{\Pi}\leq r$ for $r=0,\dots,p$. Then the set of hypotheses $H_0,\dots,H_r$ is nested,
\begin{align*}
 H_0 \subseteq H_1 \subseteq \dots \subseteq H_p,
\end{align*}
where $H_p$ corresponds to the unrestricted model. The likelihood ratio test (LRT) that compares $H_r$ and $H_p$ is applied sequentially for $r=0,1,\dots,p-1$ and continued until $H_r$ against $H_p$ cannot be rejected, and thus determine the number of cointegrating relations for $y_t$. The LRT statistic is given by
\begin{align}
	-2\log Q(H_r|H_p) = \sum_{i=r+1}^p \hat{\lambda}_i, \label{eq:LRT}
\end{align} 
where $\hat{\lambda}_1\geq\dots\geq\hat{\lambda}_p$ are the solutions to an eigenvalue problem, see \cite{johansen1996}. The asymptotic distribution of (\ref{eq:LRT}) is non-standard and previous work on estimation in high dimensional cointegrated systems discussed the problem of finding spurious cointegration relations in these when using the classic Johansen's test, see \cite{onatski2018}. The authors derive weak convergence towards the so-called Wachter distribution in high dimensional settings when the limit of the ratio between the dimensionality and number of observations exists. 

Here, we opt for a different approach by applying bootstrapping techniques as presented by \cite{cavaliere2012}, such that it is possible to infer the critical thresholds of the likelihood ratio test, numerically. Specifically, given observed data $\{y_{n}\}_{n=1}^N$, bootstrap sequences $\{y_{n}^{*(m)}\}_{n=1}^N$ for $m=1,\dots,M$ are simulated using the \emph{wild bootstrap} where the residuals are scaled by standard Gaussian distributed random variables, and for each sequence the LRT statistic $\text{LRT}^{*(m)}$ is re-computed. The empirical quantiles of $\{\text{LRT}^{*(m)}\}_{m=1}^M$ are then used for inference. 
With $r$ determined, $\hat{\beta}$ is given by the $r$ eigenvectors corresponding to $\hat{\lambda}_i,i=1,\dots,r$ and the parameter estimates $\hat{\alpha},\hat{\mu},\hat{\Sigma}$ follow by ordinary least squares estimation as described in \cite{johansen1996}.

\section{Linear Kuramoto type system}\label{sec: Kuramoto}
The classic Kuramoto model \cite{kuramoto1984} defines a system of $p$ coupled processes through the differential equations
\begin{align}
	\frac{d\theta_i}{dt} = \omega_i+\frac{\kappa}{p}\sum_{j=1}^p \sin(\theta_j-\theta_i), \for i=1,\dots,p, \label{eq: Kuramoto model}
\end{align}
where $t$ denotes continuous time. In the following, however, we consider observations of a system in discrete time. The variables $\theta_i$ are interpreted as the phases of limit cycle oscillators, with intrinsic frequencies $\omega_i$ and $\kappa$ denotes the strength of the coupling between the oscillators. By linearizing the sine function around 0 (phases synchronized) we can write the linearized Kuramoto model as
\begin{align*}
	\frac{d\theta}{dt} = \omega + \Pi\theta,
\end{align*}
where $\theta = (\theta_1,\dots,\theta_p)'$, $\omega = (\omega_1,\dots, \omega_p)'$ and $\Pi\in\RR^{p\times p}$ with the following simple structure
\begin{align}
	\Pi = \frac{\kappa}{p}\begin{pmatrix}
			1-p & 1 & \dots & 1 \\
			1 & 1-p & \dots & 1 \\
			\vdots & \vdots & \ddots & \vdots \\ 
			1 & 1 & \dots & 1-p
		\end{pmatrix} \label{eq: linearized Pi}
\end{align}
and reduced $\rank{\Pi}=p-1$. Note that this linearization only holds for $(\theta_j-\theta_i) \in [-\frac{\pi}{2},\frac{\pi}{2})$ for all $i,j=1,\dots,p$.

Consider now the following network structure inspired by the linearized version of (\ref{eq: Kuramoto model}). Let $\Pi$ denote a symmetric $p \times p$ matrix with a block structure
\begin{align}
	\Pi = \begin{pmatrix}
			\Pi_1 & 0 & \dots & 0  \\
			0 & \Pi_2 & \dots & 0 \\
			\vdots & \vdots & \ddots & \vdots  \\
			0 & 0 & \dots & \Pi_k
		\end{pmatrix}, \label{eq:Kuramoto Pi}
\end{align}
where $\Pi_i\in\RR^{p_i\times p_i}, i=1,\dots,k$ are symmetric, reduced rank matrices with $\rank{\Pi_i}=r_i=p_i-1, i=1,\dots,k$ of the form given in (\ref{eq: linearized Pi}), and $\sum_{i=1}^k p_i = p$. Hence, (\ref{eq:Kuramoto Pi}) implicitly defines a network of clusters. Note that the $p_i$'s (and hence the $r_i$'s) do not need to be identical. The network structure in (\ref{eq:Kuramoto Pi}) allows for cointegration since each $\Pi_i$ has reduced rank, and $\rank{\Pi} = \sum_{i=1}^k\rank{\Pi_i} = \sum_{i=1}^k(p_i-1)$. This implies that $\rank{\Pi}=p-k$. Interpreting this as a cointegrated system means that each cluster in (\ref{eq:Kuramoto Pi}) is driven by a stochastic $I(1)$ trend and hence the number of stochastic trends equals the number of independent clusters in the system.

The structure of $\Pi$ in (\ref{eq:Kuramoto Pi}) implies that 
\begin{align*}
	\Pi &= \diag(\Pi_1, \dots, \Pi_k) \\
	    &= \diag(\alpha_1\beta'_1, \dots, \alpha_k\beta'_k) \\
		&= \diag(\kappa_1p_1^{-1}\beta_1\delta_1\beta'_1, \dots, \kappa_1p_1^{-1}\beta_k\delta_k\beta'_k)
\end{align*}
where $\diag(\cdot)$ refers to block diagonalization. The last equation implies that $\alpha_i = \kappa_ip_i^{-1}\beta_i\delta_i$, where $\kappa_i$ is a scalar defining the coupling strength of the $i$'th cluster and $\delta_i$ is a square $r_i\times r_i$ matrix of full rank. The following Lemma states that this construction ensures that $\Pi$ is symmetric.
\begin{lemma}
If $\Pi = \Pi'$ is a symmetric $p\times p$ matrix of reduced rank $r<p$, then there exists a $p\times r$ matrix $\beta$ of rank $r$ with $\subspace(\beta)=\subspace(\Pi)$ and a symmetric $r\times r$ matrix $\delta$ of rank $r$, such that
\begin{align*}
    \Pi = \beta\delta\beta'.
\end{align*}
\end{lemma}

\emph{Proof:}  It follows that
\begin{align*}
    \Pi = V\Lambda V', V'V = VV' = I_p,
\end{align*}
where $V=(v_1, \dots, v_p)$ are the eigenvectors corresponding to the eigenvalues $\lambda_1\geq\dots\geq\lambda_p$ in $\Lambda = \diag(\lambda_1,\dots,\lambda_p)$ of which $p-r$ eigenvalues equal zero; $\lambda_{r+1}, \ldots , \lambda_p = 0$. Let $\Lambda_{(r)} = \diag(\lambda_{1},\dots, \lambda_{r})$ denote the  non-zero eigenvalues and by $V_{(r)}$ the corresponding eigenvectors, then
\begin{align*}
    \Pi = V\Lambda V'= V_{(r)}\Lambda_{(r)}V'_{(r)}, V'_{(r)}V_{(r)} = V_{(r)}V'_{(r)} = I_r.
\end{align*}
With $\beta = V_{(r)}\psi$ and $\delta = \psi^{-1}\Lambda_{(r)}\psi^{-1}$ for a symmetric $r\times r$ matrix $\psi$ of full rank $r$, the result follows. \begin{flushright} $\square$ \end{flushright}

Adding the restriction that the rows/columns of $\Pi$ must sum to 0,
% A possible construction of (\ref{eq:Kuramoto Pi}) is to choose $\alpha_i,\beta_i$ such that
% \begin{align}
% 	\alpha_i &= \begin{pmatrix*}[r]
% 				-r_ia_i & a_i & \dots & a_i \\
% 				a_i & -r_ia_i & \dots & a_i \\
% 				\vdots & \vdots & \ddots & \vdots \\
% 				a_i & a_i & \dots & -r_ia_i \\
% 				a_i & a_i & \dots & a_i
% 			\end{pmatrix*} =
% 			a_i\begin{pmatrix*}[r]
% 				-r_i& 1 & \dots & 1 \\
% 				1 & -r_i & \dots & 1 \\
% 				\vdots & \vdots & \ddots & \vdots \\
% 				1 & 1 & \dots & -r_i \\
% 				1 & 1 & \dots & 1
% 			\end{pmatrix*}	\label{eq:Kuramoto alpha i} \\
% 	\beta_i &= \begin{pmatrix*}[r]
% 				b_i & 0 & \dots & 0 \\
% 				0 & b_i & \dots & 0 \\
% 				\vdots & \vdots & \ddots & \vdots \\
% 				0 & 0 & \dots & b_i \\
% 				-b_i & -b_i & \dots & -b_i
% 			\end{pmatrix*}= 
% 			b_i\begin{pmatrix*}[r]
% 				1 & 0 & \dots & 0 \\
% 				0 & 1 & \dots & 0 \\
% 				\vdots & \vdots & \ddots & \vdots \\
% 				0 & 0 & \dots & 1 \\
% 				-1 & -1 & \dots & -1
% 			\end{pmatrix*}.	\label{eq:Kuramoto alpha i}
% \end{align}
a possible construction of (\ref{eq:Kuramoto Pi}) is to choose $\delta_i$ and $\beta_i$ such that
\begin{align}
	\delta_i &= 
			\begin{pmatrix*}[r]
				-r_i& 1 & \dots & 1 \\
				1 & -r_i & \dots & 1 \\
				\vdots & \vdots & \ddots & \vdots \\
				1 & 1 & \dots & -r_i \\
			\end{pmatrix*} \quad\text{is $r_i\times r_i$ and}	\label{eq:Kuramoto delta i} \\
	\beta_i &= \begin{pmatrix*}[r]
				1 & 0 & \dots & 0 \\
				0 & 1 & \dots & 0 \\
				\vdots & \vdots & \ddots & \vdots \\
				0 & 0 & \dots & 1 \\
				-1 & -1 & \dots & -1
			\end{pmatrix*} \quad\text{is $p_i\times r_i$},	\label{eq:Kuramoto beta i}
\end{align}
impying that
\begin{align}
	\beta_i\delta_i\beta'_i = \begin{pmatrix*}[r]
					-r_i & 1 & \dots & 1\\
					1 & -r_i & \dots & 1\\
					\vdots & \vdots  & \ddots & \vdots \\
					1 & 1 & \dots & -r_i
				\end{pmatrix*}  
				,\label{eq:Kuramoto cluster}
\end{align}
which is symmetric, rows/columns sum to 0 (since it is $r_i+1\times r_i+1)$. Furthermore, the only two parameters necessary to describe the interaction within each cluster are
\begin{enumerate}
\item $p_i=r_i+1$, the size of the cluster, and
\item $c_i=\frac{\kappa_i}{p_i}$, the strength of coupling.
\end{enumerate}
Each cluster $i$ corresponds to (\ref{eq: linearized Pi}). We will refer to a cluster of size $p_i$ as a $p_i$-cluster, hence a 2-cluster consist only of two coupled units. This definition of a cluster does not exclude 1-clusters, since this is just a single independent unit with $\delta_i = \beta_i = 0$. The construction (\ref{eq:Kuramoto cluster}) implies that the full $\Pi$ matrix can be decomposed into (sparse) $\beta,\delta$ matrices, where $\delta$ is a block diagonal matrix, i.e.,  $\Pi = \beta\delta\beta'$.

We define a \emph{linear Kuramoto type system} as a set of $k$ independent clusters of sizes $p_i, i = 1,\dots, k$, with coupling matrix on the form (\ref{eq:Kuramoto Pi}), where each cluster has a coupling strength equal to $c_i$ and a coupling structure equivalent to (\ref{eq:Kuramoto cluster}).

\section{Estimation given structural restrictions}
Under $H_r$, \cite{johansen1996} derives a likelihood ratio test for linear restrictions on $\alpha,\beta$. These are formulated as $\subspace(\alpha)\subset\subspace(A)$ and $\subspace(\beta)\subset\subspace(H)$, for given matrices $A,H$. However, in this paper we will concentrate on the structural restriction 
\begin{align}
	\Pi \in \mathcal{S}_p^r = \{M\in\RR^{p\times p} | M = M', \rank{M}\leq r \}, \label{eq: symmetric subspace} 
\end{align}
i.e., $\Pi$ must lie in $\mathcal{S}_p^r$, the subset of symmetric $p\times p$ matrices of rank smaller than or equal to $r$. We write $\mathcal{S}_p$ to denote the set of symmetric $p\times p$ matrices with no restrictions on the rank. The symmetry condition leads to a non-trivial problem of maximizing the likelihood of (\ref{eq:VECMrep}), under the two conditions: $\rank{\Pi} \leq r$ and $\Pi = \Pi'$. 

Consider first the log-likelihood function for (\ref{eq:VECMrep}) for a general (non-symmetric, possibly full rank $r\leq p$) $\Pi$, where we omit constant terms 
\begin{align}
	\log L(\Pi,\Omega) &= -\frac{N}{2}\log|\Omega|-\frac{1}{2}\sum_{n=1}^N(\Delta y_n-\Pi y_{n-1})'\Omega^{-1}(\Delta y_n - \Pi y_{n-1}).\label{eq:logLik function}
\end{align}
Defining the $p\times p$ sufficient statistics matrices $S_{ij}$, $i,j=0,1$
\begin{align}
	\begin{aligned}
		S_{00} &= \frac{1}{N}\sum_{n=1}^N \Delta y_n (\Delta y_n)' 
		\quad\quad\quad S_{01} = \frac{1}{N}\sum_{n=1}^N \Delta y_n y'_{n-1} \\
		S_{10} &= \frac{1}{N}\sum_{n=1}^N y_{n-1}(\Delta y_n)' 
		\quad\quad\quad S_{11} = \frac{1}{N}\sum_{n=1}^N y_{n-1} y'_{n-1},
	\end{aligned}\label{eq:moment matrices}
\end{align}
the log-likelihood (\ref{eq:logLik function}) can be rewritten as
\begin{align*}
	\log L(\Pi,\Omega) 	&= -\frac{N}{2}\log|\Omega|-\frac{1}{2}\tr\Bigl\{\sum_{n=1}^N(\Delta y_n - \Pi y_{n-1})(\Delta y_n - \Pi y_{n-1})'\Omega^{-1} \Bigr\} \nonumber \\
	&= -\frac{N}{2}\log|\Omega|-\frac{N}{2}\tr\Bigl\{ \bigl(S_{00}-\Pi S_{10}-S_{01}\Pi'+\Pi S_{11} \Pi'\bigr) \Omega^{-1} \Bigr\},
\end{align*}
where $\tr(\cdot)$ denotes the trace operator. The differential in the $\Omega$ direction of this expression is
\begin{align*}
    d\log L(d\Omega) &= -\frac{N}{2}\tr(\Omega^{-1}d\Omega)+\frac{N}{2}\tr(\Omega^{-1}\bigl(S_{00}-\Pi S_{10}-S_{01}\Pi'+\Pi S_{11} \Pi'\bigr) \Omega^{-1}d\Omega) \\
    &=-\frac{N}{2}\tr\Bigl(\Omega^{-1}\bigl(I_p-\bigl(S_{00}-\Pi S_{10}-S_{01}\Pi'+\Pi S_{11} \Pi'\bigr)\Omega^{-1}\bigr)d\Omega\Bigr) , %\\
    %&=-\frac{N}{2}\tr\Bigl(\Omega^{-1}\bigl(I_p-S_\Pi\Omega^{-1}\bigr)d\Omega\Bigr),
\end{align*}
which is zero for all $d\Omega$ with
\begin{align}
	\Omega_\Pi = S_\Pi = S_{00}-\Pi S_{10}-S_{01}\Pi'+\Pi S_{11} \Pi'. \label{eq: Omega function of Pi}
\end{align}
Thus, for an arbitrary fixed $\Pi$, the covariance estimator is given by (\ref{eq: Omega function of Pi}) which is explicitly dependent on $\Pi$, see also \cite{johansen1996}. Inserting (\ref{eq: Omega function of Pi}) into the log-likelihood (\ref{eq:logLik function}) yields the profile log-likelihood for $\Pi$. Up to a constant it is
\begin{align}
	\log L(\Pi) &= -\frac{N}{2}\log|\Omega_\Pi| \nonumber\\
			&= -\frac{N}{2}\log|S_{00}-\Pi S_{10}-S_{01}\Pi'+\Pi S_{11} \Pi'|	\label{eq: profile loglikelihood}
\end{align}
implying that the unrestricted maximum likelihood estimator of (\ref{eq: profile loglikelihood}) is the ordinary least squares (OLS) estimator
\begin{align}
    \hat{\Pi}_\text{ols} = S_{01}S_{11}^{-1}, \label{eq: Pi OLS estimator}    
\end{align}
which is not symmetric.

Now define $\Pi = \beta\delta\beta'$, with $\beta$ a $p\times r$ matrix of rank $r$, and $\delta$ a (possibly non-symmetric) $r\times r$ matrix of full rank $r$. The differential of (\ref{eq: profile loglikelihood}) %(\ref{eq:logLik function}) 
in the $\delta$ direction is then
\begin{align*}
    d\log L(d\delta) = -\frac{N}{2}\tr\Bigl(\Omega^{-1}\bigl(S_{01}\beta d\delta'\beta'+\beta d\delta\beta'S_{10}-\beta d\delta\beta'S_{11}\beta\delta'\beta'-\beta \delta\beta'S_{11}\beta d\delta'\beta'\bigr)\Bigr),
\end{align*}
which is zero for all $d\delta$ if 
\begin{align}
    \tr\Bigl(\Omega^{-1}\bigl(S_{01}\beta d\delta'\beta'+\beta d\delta\beta'S_{10}\bigr)\Bigr) = \tr\Bigl(\Omega^{-1}\bigl(\beta d\delta\beta'S_{11}\beta\delta'\beta'+\beta \delta\beta'S_{11}\beta d\delta'\beta'\bigr)\Bigr).\label{eq: delta score function}
\end{align}
%This leads to the following relations {\color{red}Hvad har de følgende ligheder (i rødt) med ovenstående at gøre? Og behøver vi overhovedet dem overhovedet?
% \begin{align*}
%     \tr(\Omega^{-1}\beta d\delta\beta'S_{11}\beta\delta'\beta') &= \tr(\beta'S_{11}\beta\delta'\beta'\Omega^{-1}\beta d\delta) \\
%     \tr(\Omega^{-1}\beta \delta\beta'S_{11}\beta d\delta'\beta') &= \tr(\beta'\Omega^{-1}\beta \delta\beta'S_{11}\beta d\delta') \\
%     &=\tr(\beta'S_{11}\beta\delta'\beta'\Omega^{-1}\beta d\delta).
% \end{align*}}
So far we have not imposed any symmetry conditions on $\delta$. If we assume that $\delta=\delta'$ and $d\delta= d\delta'$, we obtain the identities
\begin{align*}
    \tr\Bigl(\Omega^{-1}\bigl(S_{01}\beta d\delta'\beta'+\beta d\delta\beta'S_{10}\bigr)\Bigr) &=
    \tr(\Omega^{-1}S_{01}\beta d\delta\beta')+\tr(\Omega^{-1}\beta d\delta\beta'S_{10}) \\
    &= \tr(\beta'\Omega^{-1}S_{01}\beta d\delta)+\tr(\beta'S_{10}\Omega^{-1}\beta d\delta) \\
    &=\tr\bigl(\beta'(S_{10}\Omega^{-1}+\Omega^{-1}S_{01})\beta d\delta\bigr) 
\end{align*}
and
\begin{align*}
    \tr\Bigl(\Omega^{-1}\bigl(\beta d\delta\beta'S_{11}\beta\delta'\beta'+\beta \delta\beta'S_{11}\beta d\delta'\beta'\bigr)\Bigr) &=\tr(\Omega^{-1}\beta d\delta\beta'S_{11}\beta\delta'\beta')+\tr(\Omega^{-1}\beta \delta\beta'S_{11}\beta d\delta'\beta') \\
    &=2\tr(\beta'S_{11}\beta\delta\beta'\Omega^{-1}\beta d\delta).
\end{align*}
Thus, (\ref{eq: delta score function}) becomes
\begin{align*} 
    \tr\bigl(\beta'(S_{10}\Omega^{-1}+\Omega^{-1}S_{01})\beta d\delta\bigr)=2\tr(\beta'S_{11}\beta\delta\beta'\Omega^{-1}\beta d\delta),
\end{align*}
which holds for
\begin{align}
    \delta_\beta &= \frac{1}{2}(\beta'S_{11}\beta)^{-1}\beta'(S_{10}\Omega_{\beta,\delta}^{-1}+\Omega_{\beta,\delta}^{-1}S_{01})\beta(\beta'\Omega_{\beta,\delta}^{-1}\beta)^{-1} \label{eq: symmetric delta estimate}\\ 
    &= \frac{1}{2}(\beta'\Omega_{\beta,\delta}^{-1}\beta)^{-1}\beta'(S_{10}\Omega_{\beta,\delta}^{-1}+\Omega_{\beta,\delta}^{-1}S_{01})\beta(\beta'S_{11}\beta)^{-1}=\delta'_\beta, \nonumber
\end{align}
where the dependencies of $\delta$ on $\beta$, and of $\Omega$ on $\beta$ and $\delta$ are explicitly stated by subindices.

Since $S_{10}=S'_{01}$, (\ref{eq: symmetric delta estimate}) agrees with the result that the nearest symmetric matrix to any arbitrary quadratic matrix $M$, is the Hermitian part of $M$, $\bar{M}=\frac{1}{2}(M+M')$, in the sense that 
$||M-\bar{M}||_F \leq ||M-\tilde{M}||_F$ for any symmetric $\tilde{M}$, with equality if and only if $\bar{M}=\tilde{M}$, where $||\cdot||_F$ denotes the Frobenius norm, see \cite{fan1955}. 

Consider now
\begin{align*}
	 d \log L(d\beta) = \tr\{\Omega^{-1}(&- d\beta\delta \beta' S_{10}-\beta  \delta d\beta' S_{10} -S_{01} d \beta \delta\beta' - S_{01}\beta\delta  d \beta' \\
	&+  d \beta \delta\beta' S_{11}\beta\delta\beta' + \beta \delta d \beta' S_{11} \beta\delta\beta' + \beta\delta\beta' S_{11} d \beta\delta\beta' + \beta\delta\beta' S_{11}\beta \delta d \beta')\}.
\end{align*}
If $\beta=0$ (implying that $y_n$ is a $p$-dimensional Brownian motion and rank $r=0$) or the ordinary least squares (OLS) estimator is symmetric, and letting $\beta\delta\beta' = S_{01}S_{11}^{-1}=S_{11}^{-1}S_{10}$, the score is 0. We postulate that the set of symmetric OLS estimators is a nullset, hence the probability of this is 0. These observations imply that deriving a symmetric (low non-zero rank) estimator of $\Pi$ through the matrices $\beta,\delta$ directly from the likelihood function is not feasible. %{\color{red} Jeg synes ikke argumentet holder. Selvom det er svært at finde kan likelihooden jo godt have et minimum i mængden $\mathcal{S}_p^r$, men den kan ikke findes ved at differentiere, netop fordi mængden er så speciel (har den overhovedet et indre, så differentiering overhovedet giver mening? Det er formentlig et null set i $R^p$). Den kan vel i princippet findes ved direkte optimering. Det kan jo faktisk være at (4.9) nedenfor ER MLE'en i mængden $\mathcal{S}_p^r$. Måske man kunne udregne likelihooden i estimatoren, og så ændre, en for en, en parameter med epsilon (dvs på plads ij og plads ji), og så udregne log-likelihooden (4.5) der. Rangen kan selvfølgelig blive ændret, men den er stadig symmetrisk. Hvis alle likelihood værdier udregnet efter at have perturberet med en lille epsilon er mindre end vores estimator, er det et argument for at det er i hvert fald et lokalt maximum blandt symmetriske matricer.} 
It follows from the fact that the subset of symmetric $p\times p$ matrices in the set of $p\times p$ matrices is a null-set with respect to the Lebesgue measure on $\RR^{p\times p}$.

A possibility for estimating $\Pi$ in (\ref{eq: profile loglikelihood}) under the restriction to symmetry is to compute the Hermitian form of the ordinary least squares (OLS) estimator (\ref{eq: Pi OLS estimator})
% \begin{align}
% 	\hat{\Pi}_\text{ols} = S_{01}S_{11}^{-1}. \label{eq: Pi OLS estimator}
% \end{align}
which implies that 
\begin{align}
	\tilde{\Pi} = \frac{1}{2}(S_{01}S_{11}^{-1}+S_{11}^{-1}S_{10}) = \frac{1}{2}(\hat{\Pi}_\text{ols}+\hat{\Pi}'_\text{ols}), \label{eq: nearest symmetric OLS}
\end{align}
%Note that the Hermitian part of a $p\times p$ matrix corresponds
corresponding to the projection of $\hat{\Pi}_\text{ols}$ onto $\mathcal{S}_p$. From this, we claim that (\ref{eq: nearest symmetric OLS}) is the optimal choice of an otherwise unrestricted estimator of a symmetric $\Pi$, since $\hat{\Pi}_\text{ols}$ is the optimal unrestricted estimator for $\Pi$ and (\ref{eq: nearest symmetric OLS}) is the optimal symmetric approximation of this estimator. 

The Eckart-Young-Mirsky theorem states that the optimal approximation to a matrix by a lower rank matrix is given by the singular value decomposition (SVD)  \cite{eckart1936}. Let $M=U\Lambda V'$ denote the SVD of $M$ and assume that the entries of the diagonal $p\times p$ matrix $\Lambda = \diag(\lambda_1,\dots,\lambda_p)$ are ordered such that $\lambda_1>\lambda_2>\dots >\lambda_p$. Partition $U,V$ as $U = (U_r, U_{p-r})$ and $V = (V_r, V_{p-r})$ such that $U_r,V_r\in\RR^{p\times r}$ corresponds to the $r$ largest entries of $\Lambda$ and $U_{p-r},V_{p-r}\in\RR^{p\times (p-r)}$ corresponds to the last $p-r$ entries of $\Lambda$. Furthermore, let $\Lambda_r = \diag(\lambda_1,\dots,\lambda_r)$. Then, the optimal low rank approximation of $M$ is given by
\begin{align}
	\tilde{M} = U_r\Lambda_rV'_r. \label{eq: SVD low rank approx}
\end{align}

The SVD approximation of a low rank estimate of $\Pi$ is
\begin{align}
	\hat{\Pi}^r_\text{sym} = U_r(\tilde{\Pi})\Lambda_r(\tilde{\Pi})(V'_r(\tilde{\Pi}), \label{eq: sym Pi estimator}
\end{align}
where $U_r(\cdot),\Lambda_r(\cdot),V_r(\cdot)$ refer to the SVD matrices of rank $r$ with respect to the argument. 
 
The log-likehood is then given by (\ref{eq: profile loglikelihood}), with
\begin{align*}
	\hat{\Omega}(\hat{\Pi}^r_\text{sym}) = S_{00}-\hat{\Pi}^r_\text{sym}S_{10}-S_{01}\hat{\Pi}^{r}_\text{sym}+\hat{\Pi}^r_\text{sym}S_{11}\hat{\Pi}^{r}_\text{sym}.
\end{align*}
For the general case of a low rank approximation of a $p\times p$ matrix $\Pi$ under some additional restrictions, the problem can be stated as
\begin{align}
	\min_{M}||\Pi-M ||_F, \text{ subject to }M\in \mathcal{S}_p^r\subset\RR^{p\times p},\label{eq:structured low rank problem}
\end{align}
where $M$ is a generic matrix restricted to the subspace $\mathcal{S}_p^r$ given in (\ref{eq: symmetric subspace}).

%Under simultaneous symmetry and rank conditions, the problem becomes difficult. 
Under a strict rank condition, $\rank{M}=r$, (\ref{eq:structured low rank problem}) may not have an optimal solution in the sense that a lower rank matrix might result in a lower Frobenius norm distance \cite{chu2003}. However, the relaxed rank constraint $\rank{M}\leq r$ in $\mathcal{S}_p^r$ ensures that the Frobenius norm is minimized, given that the selected subspace is non-empty.

% This is the case for $\mathcal{S}_p^r$ which will always contain a trivial solution. Thus, the subspace $\mathcal{S}_p^r$ in (\ref{eq:structured low rank problem}) should in practice be extended to 
% \begin{align}
% 	\Pi \in \mathcal{S}_p^{r^*} = \{M\in\RR^{p\times p} | M = M', \rank{M} \leq r \}. \label{eq: extended symmetric subspace} 
% \end{align}

For a general non-empty set, $\mathcal{S}$, a low-rank approximation restricted to $\mathcal{S}$, can be solved by numerical methods such as Algorithm \ref{alg: project and lift}, which is adapted from the \emph{Lift-and-Project} algorithm in \cite{chu2003} with the order of projection and lifting switched. This is because a low rank approximation using (\ref{eq: SVD low rank approx}), of a symmetric matrix is itself symmetric. Hence, in the case where $\mathcal{S}=\mathcal{S}^{r}$ the algorithm requires only 1 iteration to complete. Note that the symmetrized low rank approximation presented above corresponds to this with target $\hat{\Pi}_\text{ols}$.
To solve (\ref{eq:structured low rank problem}) in the subspace (\ref{eq: symmetric subspace}), the algorithm can be tried for all allowed ranks, $r_1, \dots, r_n \leq r$ and the output is the matrix $M^{(i)}_j, j=1,\dots,n$ with the lowest Frobenius norm difference compared to the target matrix.

\begin{algorithm}
Initialize $M^{(0)}$ with some target $\hat{\Pi}$ of arbitrary rank, not necessarily in $\mathcal{S}$. Then for $i=1,2,\dots$ repeat until convergence between $M^{i}$ and $M^{i-1}$:
\begin{enumerate}
\item \emph{Project}: find the projection $\tilde{M}^{(i)}$ of $M^{(i-1)}$ onto $\mathcal{S}$
\item \emph{Lift}: compute the low rank approximation $M^{(i)}$ of $\tilde{M}^{(i)}$, such that $\rank{M}=r$.
\end{enumerate}
\caption{Project and Lift}\label{alg: project and lift}
\end{algorithm}

In Section \ref{sec: simulation} we compare the following estimators of $\Pi$:
\begin{enumerate}[i)]
\item The unrestricted full rank OLS estimator $\hat{\Pi}_\text{ols} = S_{01}S_{11}^{-1}$.
\item The reduced rank (non-symmetric) Johansen estimator $\hat{\Pi}_\text{J} = \hat{\alpha}\hat{\beta}'$.
%\item The reduced rank symmetric $\hat{\Pi}_{\beta,\delta}$ estimator.
\item The symmetric Johansen estimator $\hat{\Pi}_\text{proj}$ being the rank $r<p$ approximation of\\ $\frac{1}{2}(\hat{\Pi}_\text{J}+\hat{\Pi}'_\text{J})$.
\item The symmetric low rank OLS estimator $\hat{\Pi}_\text{sym}$ being the rank $r<p$ approximation of\\  $\frac{1}{2}(\hat{\Pi}_\text{ols}+\hat{\Pi}'_\text{ols})$.
\end{enumerate}
Note here that only the estimators ii)-iv) have low rank and only iii) and iv) are symmetric.

\section{High-dimensional estimation}\label{sec:HD estimation issues}
\begin{figure}[ht]
	\centering
		\includegraphics[width=0.95\linewidth]{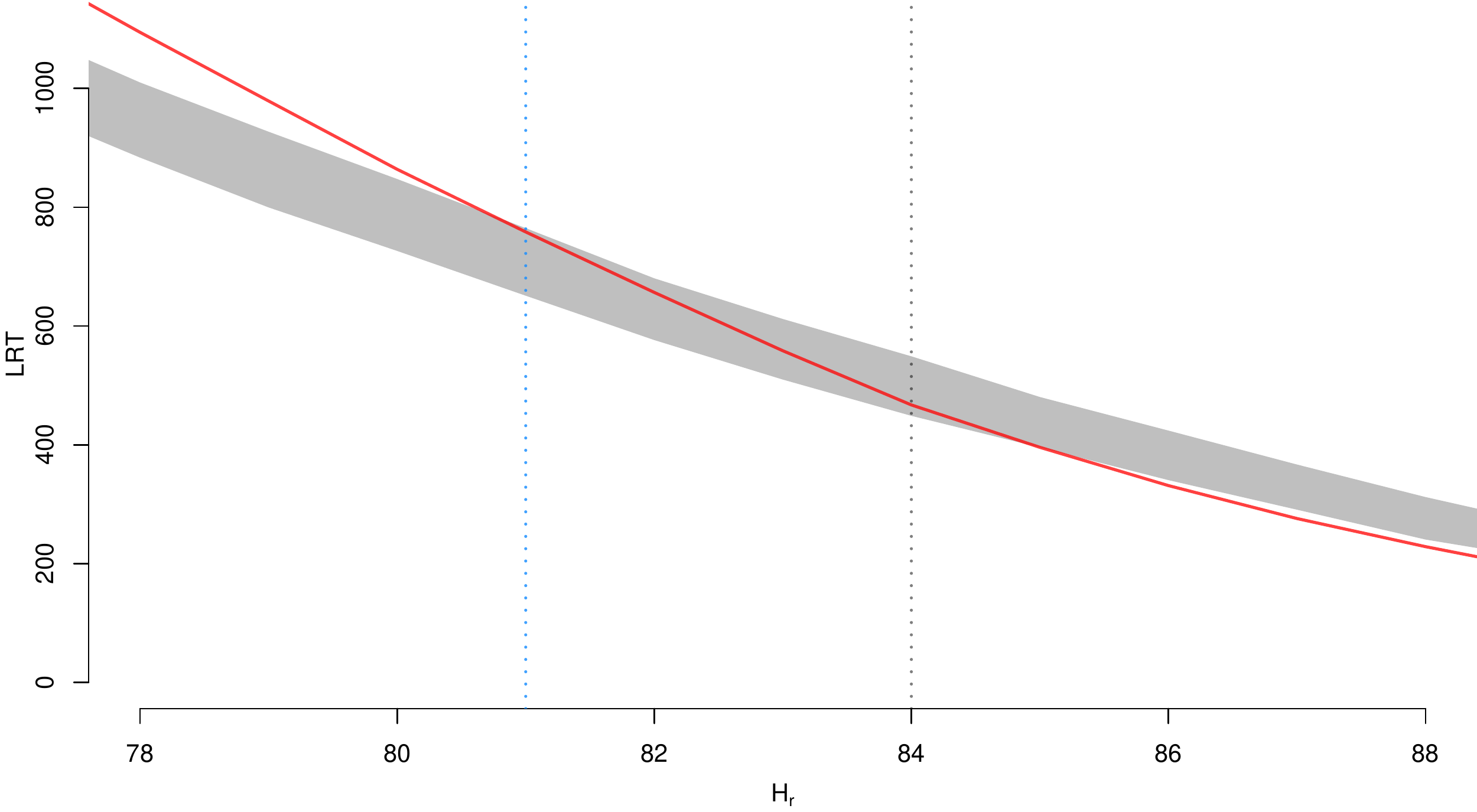}
	\caption{Example of rank estimation in a 100-dimensional linearized Kuramoto system, without any symmetry restrictions. Bootstrapped likelihood ratio test values of $H_r$ versus $H_p$ for $N=2000$. 95\% confidence bounds for the bootstrap test (gray) and the test values from the data (red). The test values enters the 95\% region at $\hat{r}=81$ (blue dashed line), whereas the true value is $r=84$ (gray dashed line). The bootstrap test is based on 300 samples. Increasing to 1000 does not change the rank estimate of $\hat{r}=81$. Increasing the number of observations $N$ will lead to more pronounced kink in the red curve at the true rank $r=84$, due to the curve being steeper for values <84. This implies a more precise estimation of the true rank.}
	\label{fig:bootstrap test}
\end{figure}

There are two crucial steps in obtaining estimates in model (\ref{eq:VECMrep}), as the rank of the system as well as $\Pi$ must be determined. When the dimension, $p$, of the model (\ref{eq:VECMrep}) increases, estimating $\hat{r}$ close to the true rank becomes progressively difficult as the rank test is prone to underestimate the rank. This is due to the construction of the test sequence, which starts with $H_r$ against $H_p$ for $r=0$. If this is rejected, then $r$ increases by 1 and this continues until $H_r$ against $H_p$ cannot be rejected or if $r+1=p$, such that $\hat{r}=p$. For some true rank $r\leq p$, this sequence is bound to reach a conclusion. However, since $H_r$ is evaluated in some distribution then it will eventually fall within a confidence region of this distribution, thus not rejecting $H_r$. This typically occurs prior to testing at the true rank $r$ as shown in Figure \ref{fig:bootstrap test}, for $N=2000$ observations of a linear Kuramoto-type system as described above (see parameter settings in Section \ref{sec: simulation}). Here the distribution of each test statistic was found by bootstrapping \cite{cavaliere2012}.

The figure demonstrates the 95\% confidence bounds (gray) for the distribution of the likelihood ratio test under $H_r$, $r=78,\dots,88$, using 300 bootstrap samples, when the true rank of the linearized Kuramoto system is 84, (see Section \ref{sec: simulation} for full model specification). The likelihood ratio test for the observed data (red) approaches the 95\% confidence area from above, and is rejected up to, and including, $H_r: r=81$. At $r=81$, the test falls within the 95\% region and therefore it is not rejected. The conclusion is therefore $\hat{r}=81$ (blue dashed line), when in fact the true rank is $r=84$ (gray dashed line). 

Inspecting Figure \ref{fig:bootstrap test} closely, the red curve displays a kink at $r=84$, indicating that at the true value there is a notable change in the likelihood ratio test trend. For a lower number of observations this is less visible, whereas for a higher number of observations, the kink is more pronounced. The numerical magnitude of the derivative of the likelihood ratio test curve, up to the true value of $r$, increases with the number of observations and thus the test is more likely to find the true value of $r$, as the trend of the red curve will cut into the grey region at a steeper angle.  Indeed, using $N=500$ observations leads to a rank estimate of 0 (indicating a random walk), $N=1000$ observations leads to a rank estimate of 10 and $N=5000$ observations estimates the rank correctly at 84. Thus, for a 100-dimensional linear Kuramoto type system, a large sample size is needed to infer the rank in an unrestricted model and thus the cointegration relations.

For the data used to produce Figure \ref{fig:bootstrap test}, the conclusion was the same when using 1000 bootstrap samples, confirming that the empirical distribution for the given number of observations is already well determined with 300 bootstrap samples.

It is of interest to asses how well the estimates of (\ref{eq:VECMrep}) model the underlying structure of the target $\Pi$ when the rank is underestimated. Intuitively we are approximating the true subspace of dimension $r$ with a subspace of dimension $\rank{\hat{\Pi}}=\hat{r}$. For $\rank{\Pi_1} = r_1\leq \rank{\Pi_2} = r_2$, we have that $\subspace(\Pi_1)\subseteq\subspace(\Pi_2)$, if the $r_1$ basis vectors for $\Pi_1$ equals the $r_1$ basis vectors of $\Pi_2$ corresponding to the $r_1$ numerically largest eigenvalues for $\Pi_2$. Thus, $\Pi_1$'s basis vectors are the $r_1$ most important basis vectors of the $\Pi_2$ space. Therefore, underestimating the rank of $\Pi$ implies that the basis vectors of the estimation space will be linear combinations of the (most important) basis vectors of the true $\Pi$-space.

The quality of the estimation depends on how close the estimation rank is to the true rank. This can be visualized by defining a measure of "closeness" for two subspaces as defined by the matrices $\Pi$ (true subspace) and $\hat{\Pi}$ (estimated subspace). For this purpose, one can use a generalized version of the vector angle, using the Frobenius inner product $\langle U,V\rangle_F = \tr(U'V)$ for matrices $U,V\in\RR^{p\times p}$. The angle between the matrices $U$ and $V$ is defined as
\begin{align}
	\Theta(U,V) = \arccos\Biggl(\frac{\langle U,V\rangle_F}{\sqrt{\langle U,U\rangle_F\langle V,V\rangle_F}}\Biggr).\label{eq:Matrix angle}
\end{align}
For $U=0$ and/or $V=0$ we define $\Theta(U,V) = \pi/2$, i.e., the 0 matrix is always orthogonal to itself and any other. 
\begin{figure}[h]
	\centering
	\includegraphics[width=0.95\linewidth]{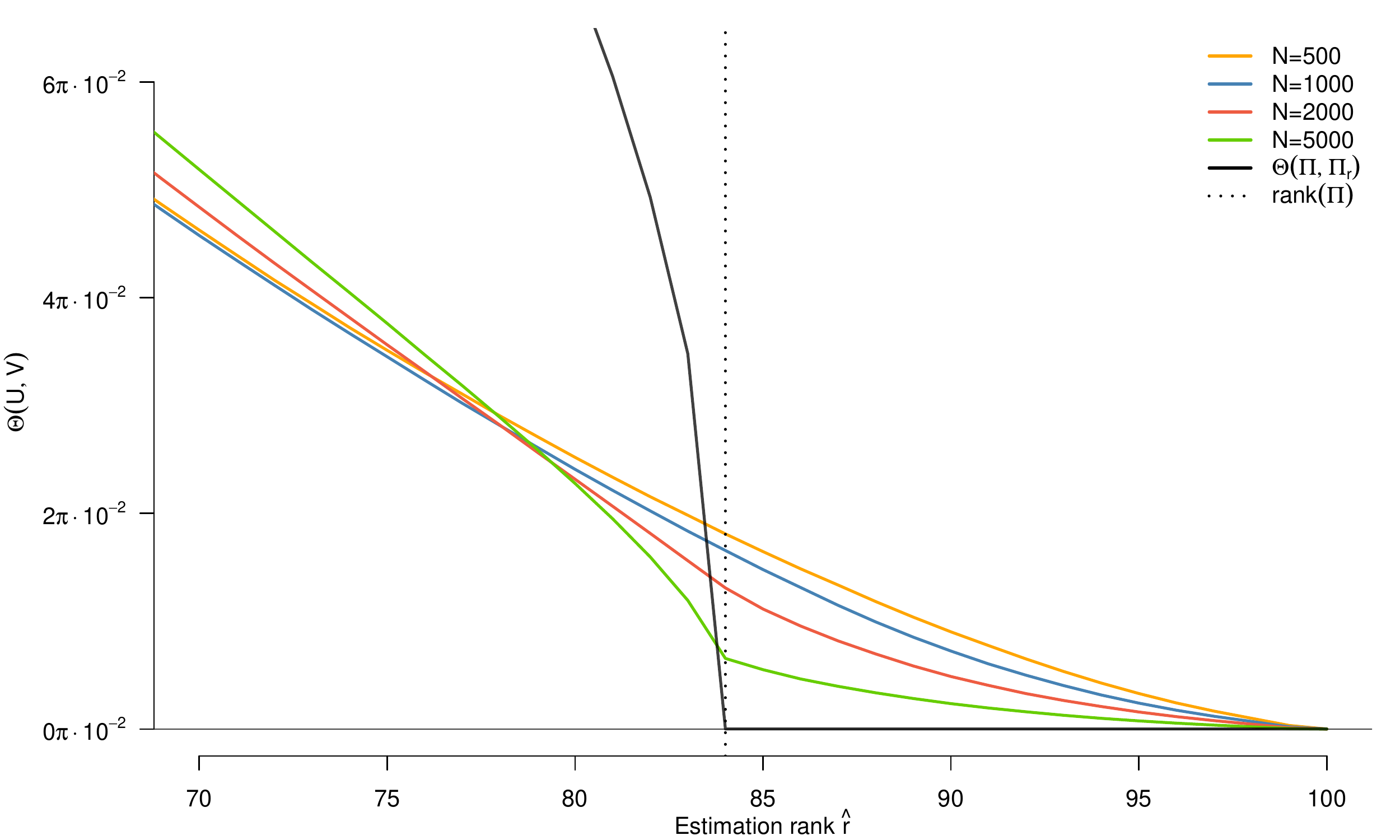}
	\caption{Matrix angles $\Theta(\hat{\alpha}_r\hat{\beta}_r',\hat{\Pi}_\text{ols})$ between the Johansen estimator $\hat{\alpha}_r\hat{\beta}_r'$ and the OLS estimator $\hat{\Pi}_\text{ols}$, for a 100-dimensional system, when estimating $\alpha$ and $\beta$ under varying rank and number of observations $N=500,1000,2000,5000$. Shown are median angles, based on 500 simulations for each combination of rank $\hat{r}$ and observations $N$. The reference (black) $\Theta(\Pi^{84},\Pi^r)$ shows the angle between the true simulation matrix $\Pi$ and low rank approximations of this using (\ref{eq: SVD low rank approx}). The vertical dashed line marks $\rank{\Pi^{84}}$. As $\hat{r}$ decreases towards 0, all angles increase monotonically to $\pi/2$.}
	\label{fig:Angle of Pi estimation}
\end{figure}
Figure \ref{fig:Angle of Pi estimation} displays the matrix angles between the Johansen estimator (no symmetry restriction) $\hat{\Pi}_\text{J}$ for $\hat{r}=70,\dots,100$ against the (unrestricted) full rank OLS estimator $\hat{\Pi}_\text{ols}$, for a varying number of observations $N=500,1000,2000,5000$ of the same Kuramoto system simulated in Figure \ref{fig:bootstrap test} (for parameter settings, see Section \ref{sec: simulation}). Shown are median angles, based on 500 replications for each combination of observations $N$ and rank $r$. As a reference, the angles between the low rank approximations $\Pi^r$ of the true simulation matrix $\Pi^{84}$ are presented as well. The vertical dashed line marks $\rank{\Pi}=84$.
 
The reference angles show that as the rank increases, the low rank approximations become better, however when reaching the true rank of $\Pi$, then $\Theta(\Pi,\Pi_r)=0$ for $r\geq84$. In contrast, the angles between the Johansen estimator and the OLS estimator decrease past the true rank of $\Pi$, due to the fact that $\rank{\hat{\Pi}_\text{ols}}=100$. However, as the number of observations increase, the kink noticeable for the reference angles $\Theta(\Pi^{84},\Pi^r)$ at $\hat{r}=84$, starts to appear, suggesting that as the number of observations increase, the OLS estimator is well approximated at the rank of the true simulation matrix $\Pi$. 

Generally, as $\hat{r}$ increases, the matrix angles decrease from orthogonality ($\pi/2$) to a minimum. Underestimating the rank will influence the estimation negatively, but the severity is dependent on how far from the true rank $\hat{r}$ is. Overestimation ensures a lower matrix angle, but impacts estimation negatively due to an excess number of parameters to be estimated. From Figure \ref{fig:Angle of Pi estimation} it is evident that estimating in the vicinity of the true rank is optimal when fitting the model, however it is not of crucial importance to find the exact true rank and we can obtain fair estimates of $\Pi$ when the rank is underestimated by a relatively small amount. 
								
\section{Simulation}\label{sec: simulation}
\begin{figure}[ht]
	\centering
	\includegraphics[width=0.95\linewidth]{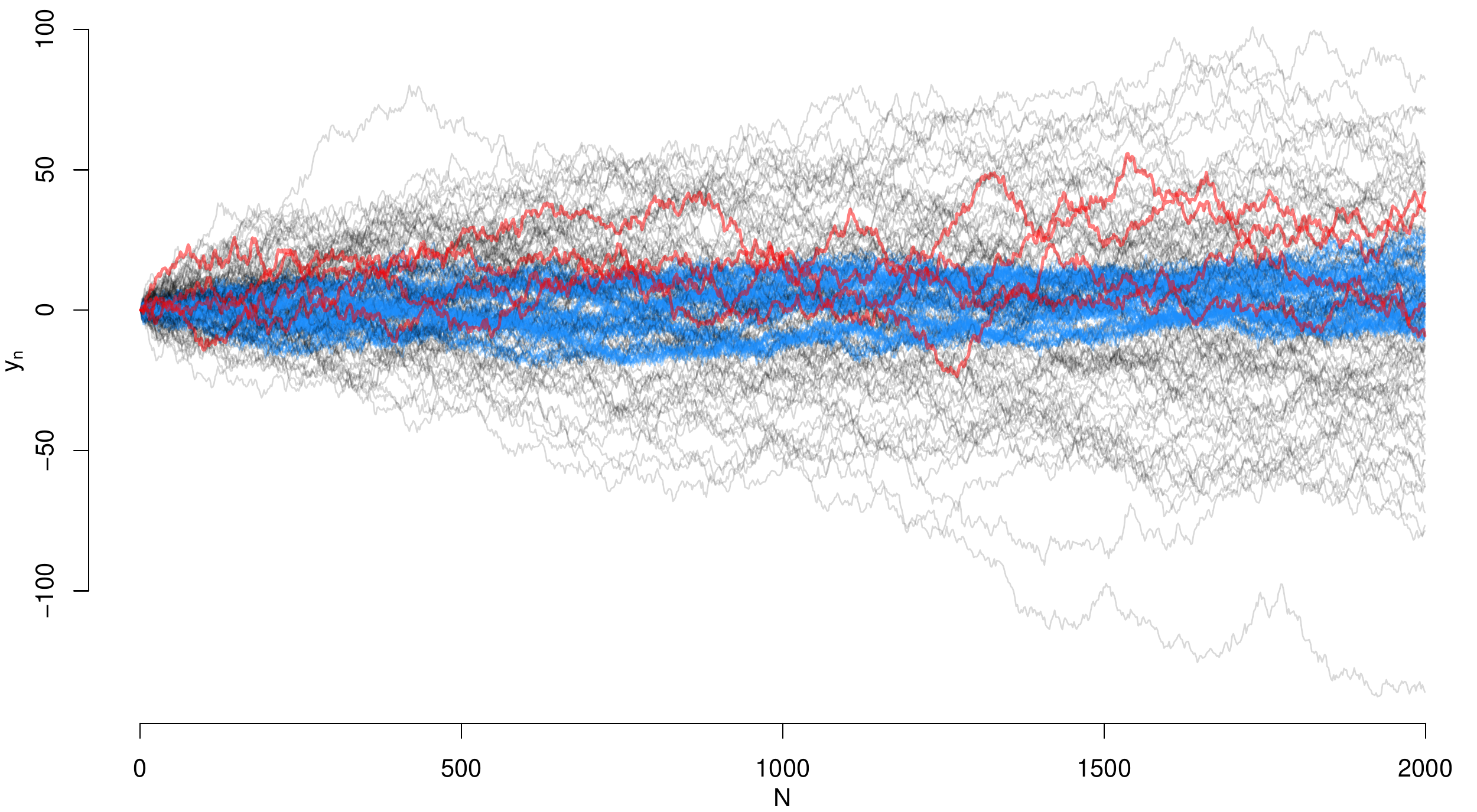}
	\caption{Simulation of a linear Kuramoto system with 12 8-clusters (blue) and 4 1-clusters (red). A 100 dimensional multivariate random walk (black) is superimposed for reference. The 1-clusters (red) could essentially be any of the random walk processes due to the diagonal covariance structure of the cointegrated system.}
	\label{fig:Kuramoto System}
\end{figure}

Consider a system with $\Pi$ as in (\ref{eq: linearized Pi}-\ref{eq:Kuramoto Pi}) and $p=100$ with twelve 8-clusters and four 1-clusters, i.e., a system of 12 coupled clusters each of size 8 and 4 independent processes. The order of the simulated processes is scrambled before any estimation takes place, such that adjacent processes of the observations are not necessarily part of the same cluster. For presentation purposes, the clusters are reordered correctly when visually displaying results. We let each cluster be coupled with varying strengths $c_i=a_ib_i \in [0.5; 2], i=1,\dots,12$ and assume that $\Omega = I_p$. For the 4 independent processes, the coupling strengths are implicitly set to 0, cf. Section \ref{sec: Kuramoto}. The chosen diagonal covariance matrix implies that if two processes do not cointegrate, then they are independent. However, this is not generally true. If the covariance is non-zero between to processes that do not cointegrate, they still interact through the covariance. A simulation of $N=2000$ observations with initial condition $y_0=0$ is presented in Figure \ref{fig:Kuramoto System}, where the coloring is for easy identification of the 8-clusters (blue), the 1-clusters (red) and a pure 100-dimensional random walk for reference. 
It is noticeable how each cluster, as a whole, behaves like a random walk. The  individual processes within each cluster overlap, appearing as a few blue processes represented by thick lines, visible in Figure \ref{fig:Kuramoto System}. The coupling strength is so strong that the internal cluster processes never stray far from the cluster, and as a whole they behave as a stationary process. The random walk reference shows how tightly the cointegrated processes are knitted.

\subsection{Estimation of $\Pi$}
The rank estimation of $\Pi$ was performed with bootstrapping. From 300 bootstrap samples, the rank was determined as $\hat{r}=81$, see Figure \ref{fig:bootstrap test}. Given this rank estimate, we compare 4 estimators for $\Pi$: i) the unrestricted OLS estimator $\hat{\Pi}_\text{ols}$, ii) the (unrestricted) Johansen estimator $\hat{\Pi}_\text{J}$, iii) the symmetrized Johansen estimator $\hat{\Pi}_\text{proj}$ and iv) the symmetrized low rank OLS estimator $\hat{\Pi}_\text{sym}$.

\begin{figure}[h]
	\centering
	\includegraphics[width=0.95\linewidth]{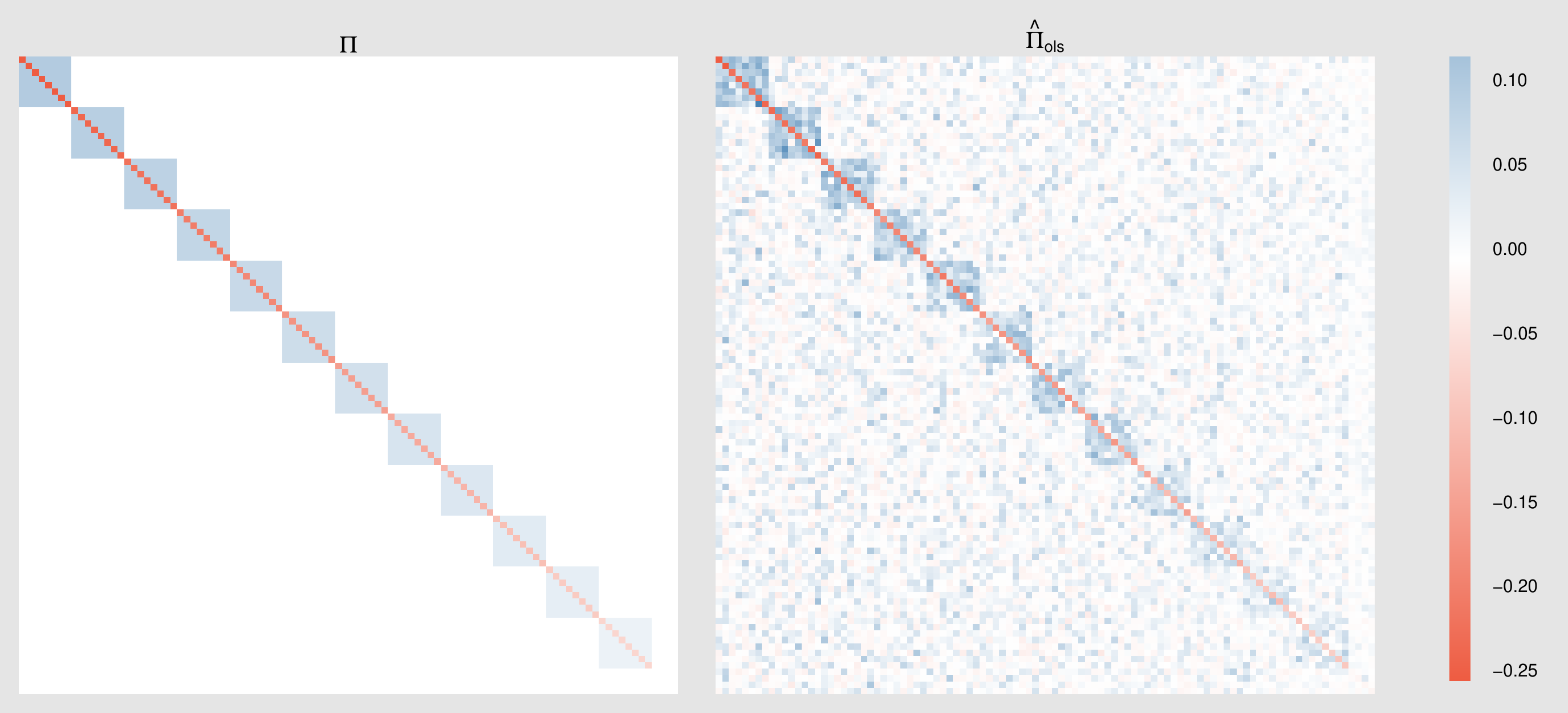}
	\caption{True $\Pi$ used for simulation (left side), estimated $\hat{\Pi}_\text{ols}$ by OLS (right side). Notice the decreasing coupling strengths in the true matrix indicated by lighter blue colors. The lower right corner of $\Pi$ is 0. The entries of the matrices are ordered correctly for easy comparison.}
	\label{fig:Estimating Pi OLS}
\end{figure}

Figure \ref{fig:Estimating Pi OLS} presents the true simulation matrix $\Pi$ with the unrestricted (full rank) OLS estimator $\hat{\Pi}_\text{ols}$, correctly ordered. We detail the ordering in Section \ref{sec: simulation}.3 below.
The decreasing coupling strength of the Kuramoto clusters is evident in both matrices, indicated by the fading colors across the block diagonal. Note the lower right corner, where the 4 1-clusters are represented as 0 entries in $\Pi$. The OLS is capable of reproducing the structure of the true $\Pi$ to some extend.

\begin{figure}
	\centering
	\includegraphics[width=0.99\linewidth]{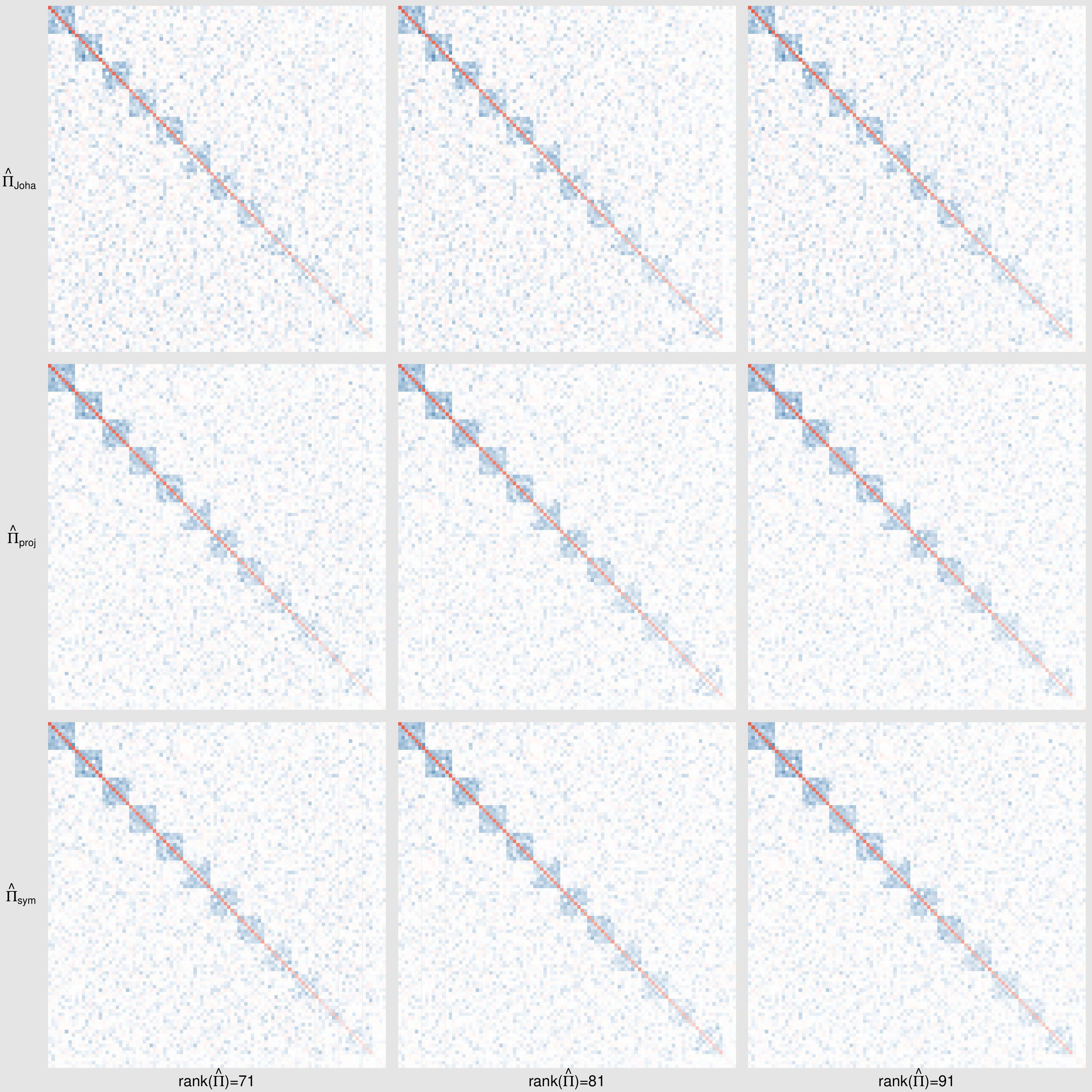}
	\caption{Estimating $\Pi$ with the Johansen estimator $\hat{\Pi}_\text{J}$, the symmetrized Johansen estimator $\hat{\Pi}_\text{proj}$ and the symmetrized low rank OLS estimator $\hat{\Pi}_\text{sym}$, under various low rank approximations for $\hat{r}=71,81,91$, when the estimated rank is 81. The middle column corresponds to the estimates under the inferred rank, the left column corresponds to estimates under an (extreme) low rank and the right column corresponds to estimates under an (extreme) high rank. For $\hat{r}=71$, estimates are more noisy, but for $\hat{r}=81, 91$, visual comparisons are similar across rows. Also, $\hat{\Pi}_\text{proj}$ and $\hat{\Pi}_\text{sym}$ look very similar. All matrices are ordered correctly for easy comparison.}
	\label{fig:Pi est with ranks}
\end{figure}
Figure \ref{fig:Pi est with ranks} shows the 3 types of low rank estimator considered, estimators ii)-iv) under the inferred rank $\hat{r}=81$ as well as extreme under/over-estimated ranks $\hat{r}=71,91$. Evidently, parts of the structure of $\Pi$ is recovered in all cases, but for the underestimated rank, the plots are slightly more noisy. However, comparing estimators of rank 81 and 91, there are no visual differences. Comparing the symmetric estimators $\hat{\Pi}_\text{proj}$ and $\hat{\Pi}_\text{sym}$, these are indistinguishable visually, whereas comparing them to the (non-symmetric) $\hat{\Pi}_\text{J}$, minor visual differences appear.

\subsection{Comparing estimators under structural restrictions}
We compare the four estimators with respect to two measures: 1) the matrix angle between the estimator and the true matrix $\Pi$ and 2) the standard deviation of the entries of the estimator that lies outside the true block diagonal structure, i.e. the indices that are 0 in the true matrix $\Pi$. The matrix angle measures how well the estimator captures the true structure of the simulation matrix $\Pi$ whereas the standard deviation measure describes the amount of noise in the estimation of entries that should optimally be 0. The individual estimates are also compared to the measures for $\hat{\Pi}_\text{ols}$ which is non-symmetric and of full rank. 
\begin{table}[ht]
\centering
\begin{tabular}{l rrr | rrr}
  \toprule
	& \multicolumn{3}{c}{Matrix angles $\Theta(\hat{\Pi},\Pi)$} & \multicolumn{3}{c}{std. dev. of 0 entries}\\
	\midrule
 	& $\hat{r}=71$ & $\hat{r}=81$ & $\hat{r}=91$ & $\hat{r}=71$ & $\hat{r}=81$ & $\hat{r}=91$ \\ 
  \midrule
	$\hat{\Pi}_\text{ols}$ & & 0.8492 & & & 0.0176 & \\
	$\hat{\Pi}_\text{J}$ & 0.9018 & 0.8678 & 0.8514 & 0.0184 & 0.0180 & 0.0176 \\  
  	$\hat{\Pi}_\text{proj}$ & 0.8249 & 0.7899 & 0.7861 & 0.0149 & 0.0142 & 0.0142 \\
	$\hat{\Pi}_\text{sym}$ & 0.8230 & 0.7891 & 0.7847 & 0.0148 & 0.0142 & 0.0142 \\ 
  \bottomrule
\end{tabular}

\caption{Measures of estimators of $\Pi$. Matrix angles $\Theta(\hat{\Pi},\Pi)$ measures how the estimator captures the structure of $\Pi$ and the standard deviation measures the amount of noise in the estimator.} 
\label{tab: estimator matrix measures}
\end{table}

Table \ref{tab: estimator matrix measures} presents the results under the inferred rank $\hat{r}=81$ as well as the under/over-estimated ranks $\hat{r}=71,91$. It is clear that the symmetric estimators $\hat{\Pi}_\text{proj}$ and $\hat{\Pi}_\text{sym}$ are superior in capturing the dynamics as well as lowering the noise in (true) zero entries. Comparing the angles for $\hat{\Pi}_\text{proj}$ and $\hat{\Pi}_\text{sym}$ across ranks, we see that the decrease from $\hat{r}=71$ to $\hat{r}=81$ is much larger than from $\hat{r}=81$ to $\hat{r}=91$, suggesting a lower marginal gain. The standard deviation measure does not improve with increasing rank for the symmetric estimators.
\begin{figure}[h]
	\centering
	\includegraphics[width=0.95\linewidth]{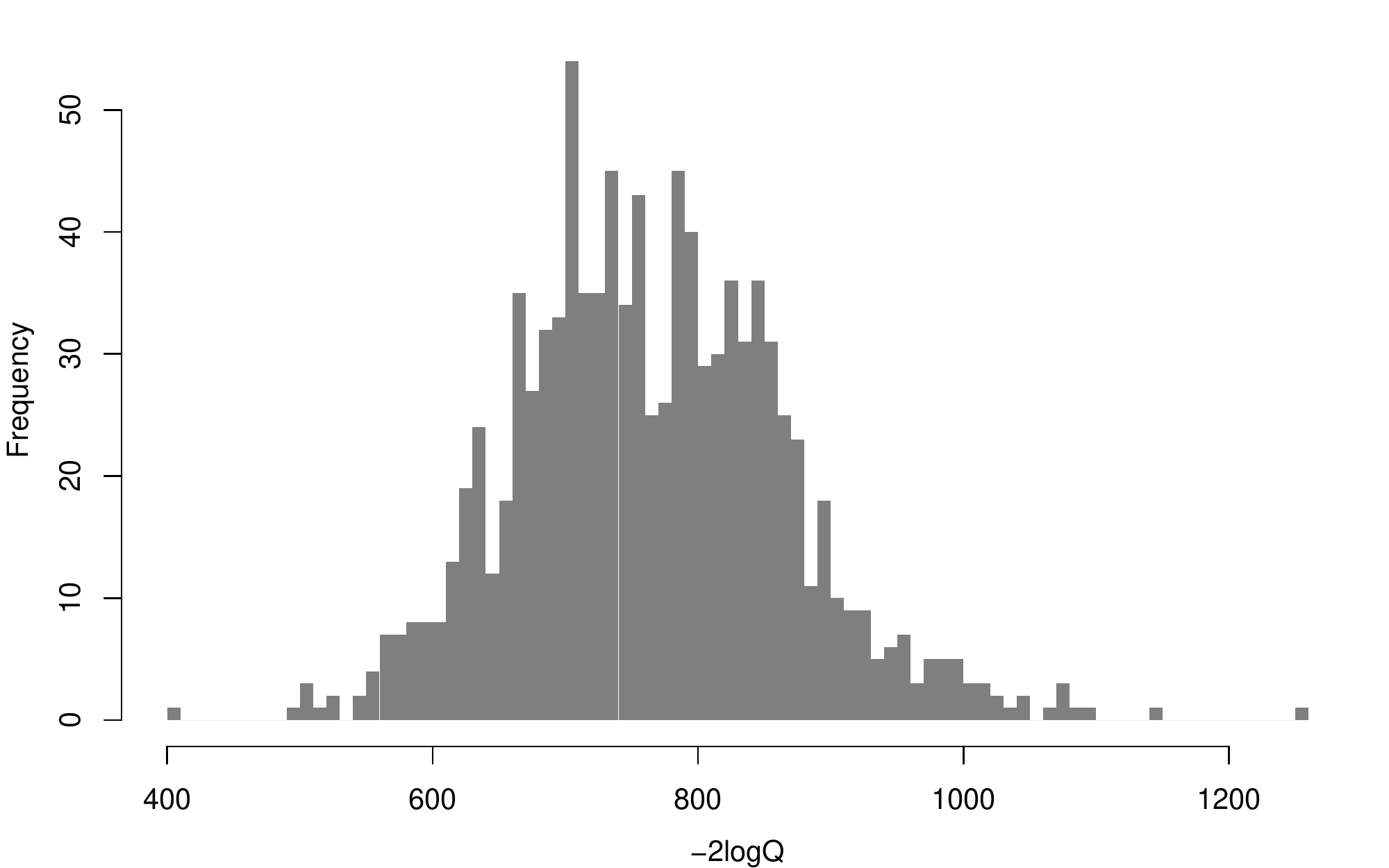}
	\caption{Comparing the LRT values of the $\hat{\Pi}_\text{proj}$ and the $\hat{\Pi}_\text{sym}$ estimators: $-2\log Q(\hat{\Pi}_\text{proj},\hat{\Pi})-\bigl(-2\log Q(\hat{\Pi}_\text{sym},\hat{\Pi})\bigr)$, based on 1000 replications. All differences are positive, implying that the LRT values of $\hat{\Pi}_\text{sym}$ are lower in all simulations, further implying that this estimator is the superior of the two.}
	\label{fig: lrt difference of proj and sym}
\end{figure}

Comparing LRT values at $\hat{r}=81$ for the symmetric estimators, the $\hat{\Pi}_\text{sym}$ has a smaller LRT value of 27769 %27768.63 
against a value of 28410 %28410.49 
for the $\hat{\Pi}_\text{proj}$ estimator. Indeed simulating this difference 1000 times, we find that this is generally the case. Figure \ref{fig: lrt difference of proj and sym} shows a histogram of the difference of the two LRT values $-2\log Q(\hat{\Pi}_\text{proj},\hat{\Pi})-\bigl(-2\log Q(\hat{\Pi}_\text{sym},\hat{\Pi})\bigr)$. Ranking the two similar symmetric low rank estimators, the $\hat{\Pi}_\text{sym}$ comes out on top with slightly better performance than $\hat{\Pi}_\text{proj}$ based on the measures considered here and it comes with a higher likelihood value. This is of course a result of approximating the unrestricted OLS estimator with a symmetric low rank version as opposed to approximating the Johansen estimator, which is already under a rank condition when being estimated. On the other hand, it is the Johansen estimator on which the rank test is based, or in the case where $r=p$ the OLS estimator, as they coincide. Hence, both are necessary for the inference considered here, but in conclusion the approximation of the OLS is superior to determine the underlying structure of the high-dimensional process.

\subsection{Recovering clusters from a random ordering}
Estimation of $\Pi$ does not rely on the individual processes being ordered according to the cluster structure used in the simulation. However, the observations cannot be assumed to follow this ordering, and we demonstrate how to recover the structure from an estimate of $\Pi$ using the CNM algorithm from \cite{clauset2004}. The algorithm is a graphical approach to group nodes in a graph with weighted edges by maximizing a modularity score measuring the community structure in a network, where a measure of 0 corresponds to complete randomness in the network structure. Using the modularity score, the CNM algorithm determines the number of clusters intrinsically, thereby only requiring a matrix of weighted edge connections as input. Here we used $\abs(\hat{\Pi}_\text{sym})$ as weights, where $\abs(\cdot)$ denote the entrywise absolute value, since it is the magnitude of entries that determine the connection strength, the sign being irrelevant. 
\begin{table}[ht]
\centering
\addtolength{\leftskip} {-2cm} % increase (absolute) value if needed
\addtolength{\rightskip}{-2cm}
\begin{footnotesize}
\begin{tabular}{r l}	
	\hline
        Coupling & True Constituents \\
 	 \hline	
2.00& \{15, 33, 54, 60, 62, 90, 93, 94\} \\
1.86& \{9, 34, 42, 53, 73, 75, 88, 92\} \\ 
1.73 & \{4, 23, 24, 30, 50, 56, 76, 100\} \\ 
1.59& \{12, 13, 46, 48, 57, 68, 72, 80\} \\ 	             
1.45 & \{3, 19, 22, 40, 63, 65, 74, 82\} \\ 
1.32 & \{1, 7, 11, 25, 41, 69, 71, 79\} \\ 
1.18 & \{5, 21, 39, 43, 47, 49, 59, 95\} \\ 
1.05& \{8, 28, 31, 36, 58, 61, 89, 97\} \\ 
0.91 & \{20, 27, \miss{45}, 51, 55, 78, 81, \miss{98}\} \\ 
0.77& \{6, 26, 37, 38, 52, 64, 83, 99\} \\ 
0.64 & \{10, \miss{14}, \miss{16}, 32, \miss{35}, 44, \miss{85}, \miss{96}\} \\   
0.50 & \{18, 29, 66, 67, 70, 86, 87, 91\} \\ 
0 & \{\miss{2}\} \\ 
0 & \{\miss{17}\} \\ 
0 & \{\miss{77}\} \\ 
0 & \{\miss{84}\} \\  
          \hline
\end{tabular}
          \hspace*{10px}
\begin{tabular}{l}
  \hline
	Estimated Constituents\\ 
          \hline
\{15, 33, 54, 60, 62, \miss{77}, 90, 93, 94\} \\ 
\{9, 34, 42, 53, 73, 75, 88, 92\} \\ 
\{4, 23, 24, 30, 50, 56, 76, 100\} \\ 
\{12, 13, 46, 48, 57, 68, 72, 80\} \\
\{3, 19, 22, 40, 63, 65, 74, 82\} \\ 
\{1, 7, 11, 25, 41, 69, 71, 79\} \\
\{5, 21, 39, 43, 47, 49, 59, 95\} \\ 
\{8, 28, 31, 36, 58, 61, 89, 97\} \\ 
\{\miss{14}, \miss{17}, 20, 27, 51, 55, 78, 81, \miss{85}\} \\ 
\{\miss{2}, 6, 26, 37, 38, 52, 64, 83, \miss{84}, 99\} \\                                 
\{10, 32, 44\} \\
\{\miss{16}, 18, 29, \miss{35}, \miss{45}, 66, 67, 70, 86, 87, 91, \miss{96}, \miss{98}\} \\ 
 \\ \\ \\ \\
          \hline
\end{tabular}
\end{footnotesize}
	        \caption{Processes in each cluster ordered by coupling strength compared to estimated clusters. Left: the true coupling strengths and constituents of each of the 16 clusters. Right: Estimated constituents, using the CNM algorithm. Processes on one side, but not the other appear in red. Evidently as the coupling strength decreases more classification errors occurs. Furthermore all four independent processes are misplaced in other clusters. The CNM algorithm uses a modularity score to determine the number of clusters.\label{tab: recovered clusters}}
\end{table}

\begin{figure}[ht]
	\centering
	\includegraphics[width=0.95\linewidth]{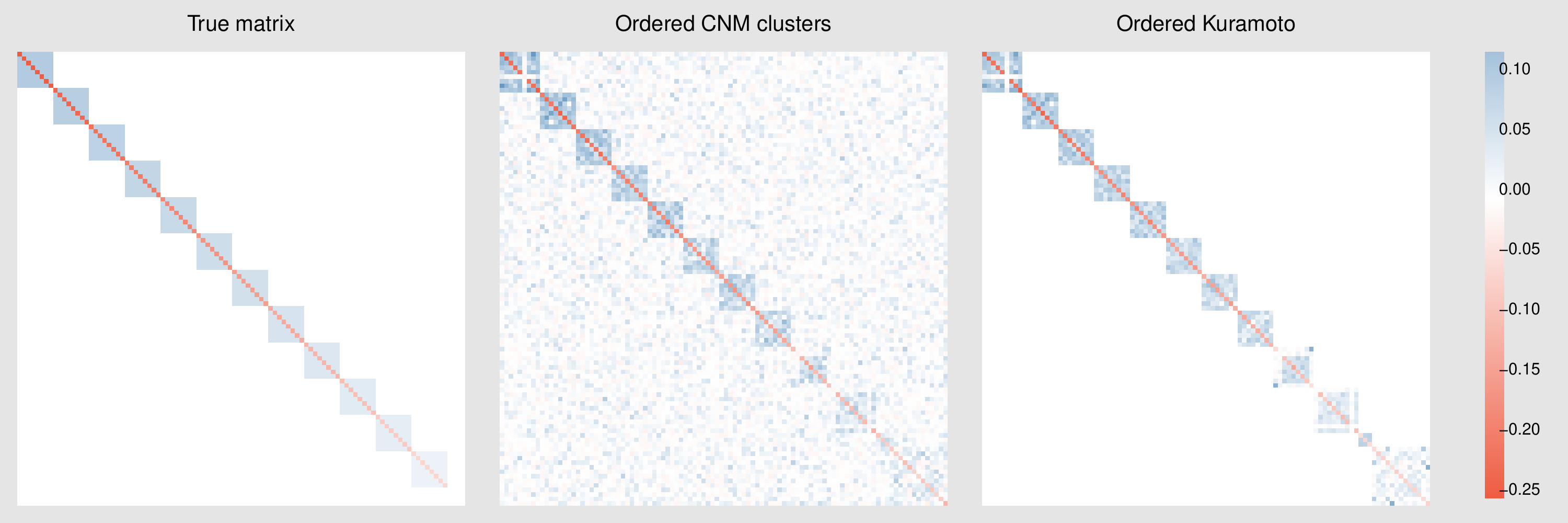}
	\caption{Estimate of $\Pi$ with $\hat{r}=81$. Left: True matrix. Middle: clusters estimated by the CNM algorithm, ordered by average cluster coupling strength as in Table \ref{tab: recovered clusters}. A misplaced independent 1-cluster is visible in the upper left corner. Right: Kuramoto matrix estimated using the same cluster estimates used in the middle. Clusters are assumed independent, hence the individual components are glued together from individual estimates for visual comparison.}
	\label{fig: recovered clusters}
\end{figure}

Table \ref{tab: recovered clusters} shows the output from the algorithm against the true cluster structure ordered by decreasing coupling strength. There are 12 8-clusters of varying coupling strengths $c_i, i=1,\dots,12$ and 4 1-clusters in the true structure (the coupling strength is zero for the last four processes). The CNM algorithm identifies the 8-clusters closely for $c_i>1$, missing only 1 process (77) which is one of the 4 independent 1-clusters. When the coupling strength decrease to $c_i<1$, more classification errors appear. Especially the two clusters with the lowest $c_i$'s are mixed up and all independent 1-clusters are misplaced in larger clusters. Figure \ref{fig: recovered clusters} (middle) displays the $\hat{\Pi}$ matrix ordered by the cluster structure identified by the CNM algorithm compared with the true matrix (left side). The misplaced independent 1-cluster in the strongest coupled cluster is visible in the upper left corner. The maximum modularity is estimated at 0.357 by the CNM algorithm, which is indicative of significant community structure according to \cite{clauset2004}. 

Figure \ref{fig: recovered clusters} (right) shows an estimate of the Kuramoto structure, see Section \ref{sec: Kuramoto}. Assuming independence among clusters and using the CNM clusters as input, we estimate the individual cluster's internal coupling structure separately using the $\hat{\Pi}_\text{sym}$ estimator for each of the partitions of the full system. This results in a structure similar to that of the middle of Figure \ref{fig: recovered clusters}, but without the noise off the block diagonal, as these entries are set to zero by construction. Therefore, the individual cluster structure becomes more pronounced. The misplaced independent process in the strongest coupled cluster in the upper left corner is clearly visible.

\section{Discussion}
This paper discussed estimation of a cointegrated system in a high-dimensional setting such as a symmetric linearized Kuramoto system. While previous work has derived asymptotic properties of high-dimensional cointegrated systems, see \cite{onatski2018}, we opted instead for bootstrapping for rank estimation, as described in \cite{cavaliere2012}. Bootstrapping is simple to implement and performs well as demonstrated in a simulation example.
 
The claim that the sequential testing of the rank is prone to underestimating of r, was supported by a visual argument on how the likelihood ratio test behaves when associated with a distribution to evaluate the test statistic. However, it was also demonstrated by examples, that this underestimation is not a major cause for concern, when the dimension of the system is high and the rank is underestimated to a lesser extend, contrary to low dimensional systems where determining the precise rank is an essential task. This is also supported by the findings of \cite{holberg2022} where the exact distributions of the estimators of the cointegration matrix $\Pi$ under misspecification of the rank are found. Underestimation of the rank introduces a bias, however, it is usually small and compensated by a smaller variance. Overestimation of the rank does not introduce bias but it inflates the variance.
 
The underestimation of the rank by sequential testing seems contradictory to the findings of \cite{onatski2018}, where they showed that the rank is asymptotically overestimated, leading to wrongly inferred cointegration relations. However, the asymptotic regime investigated in \cite{onatski2018} is different from ours. They consider a high-dimensional setting where both $p$ and $N$ goes to infinity, keeping the proportion $p/N$ constant. We do not consider an asymptotic regime, but finite sample size and large but fixed $p$, say $p \gg 10$.
 
Determining the rank of a cointegrated system is important in order to find the
correct number of stochastic trends that drive the system, however when $p\to\infty$, the number of
sequential tests can become a problem. Therefore, it might be of value to investigate alternatives
in order to minimize the error of selecting a rank very far from the true value, see \cite{levakova2022} for a review of penalization methods in cointegration models including evaluation and comparisons between methods for different error measures.

%This paper discussed estimation of a cointegrated system in a high-dimensional setting such as a symmetric linearized Kuramoto system. While previous work on high dimensional cointegration has derived asymptotic properties of high-dimensional cointegrated systems, see \cite{onatski2018}, we opted instead for bootstrapping for rank estimation, as described in \cite{cavaliere2012}. Bootstrapping is simple to implement and performs well as demonstrated in a simulation example. The claim that the sequential testing of the rank is prone to underestimating of $r$, was supported by a visual argument on how the likelihood ratio test behaves when associated with a distribution to evaluate the test statistic. However, it was also demonstrated by examples, that this underestimation is not a major cause for concern, when the dimension of the system is high and the rank is underestimated to a lesser extend, contrary to low dimensional systems where determining the precise rank is an essential task. Determining the rank of a cointegrated system is important in order to find the correct number of stochastic trends that drive the system, however when $p\to\infty$, the number of sequential tests can become a problem. Therefore, it might be of value to investigate alternatives in order to minimize the error of selecting a rank very far from the true value.

The task of estimating the matrix $\Pi$ under the non-standard restriction to symmetry, as well as the usual low rank condition, required a two step procedure that separated the tasks. We presented an estimator, based on the standard OLS estimator which was sequentially approximated by a symmetric and low rank matrix. This estimator proved superior to the standard low rank Johansen estimator as well as a symmetrized version of this, measured by the matrix angle to the true matrix and the difference in the likelihood ratio test statistic between the two symmetrized estimators. Based on the symmetric low rank estimator, we used the graphical CNM algorithm to recover the underlying cluster structure of the simulated linear Kuramoto system. This proved to be effective for stronger coupling strengths, recovering most of the clusters. However, a few of the weaker coupled clustered were mixed up. The few independent random walks added to the Kuramoto system were all placed in some of the CNM estimated clusters, although all of these should have been placed in separate clusters of the single processes. %Visualizing the estimated Kuramoto matrix, one of these misplacements in a strongly coupled cluster was especially conspicuous. 

The CNM algorithm was selected based on it's simplicity and relatively strong performance in recovering the Kuramoto clusters. However, other (graphical) algorithms might provide equally good or better results. The CNM estimates also the number of clusters, hence an algorithm assuming a fixed number might be better suited. From the estimated rank $\hat{r}$ of a $p\times p$ matrix $\Pi$, the number of clusters in a linearized Kuramoto system is $p-\hat{r}$. Thus, when the rank is underestimated, the number of clusters are overestimated. This knowledge could be taken into account as well, such as the maximum number of clusters to search for.

\section{Conclusion}
We estimated the rank $r$ of a high-dimensional cointegrated system, by bootstrapping a distribution for the likelihood ratio test. Although the rank is underestimated, we showed that this is not a major concern in a high-dimensional setting, when the underestimation is relatively small. We also derived an estimator that restricts the matrix of the system to be both symmetric and of low rank. This estimator is based on the usual OLS estimator, using a two-step procedure to impose the conditions of symmetry and low rank. We also defined and analyzed a simulation of a 100-dimensional linearized Kuramoto system and used a graphical approach to recover the underlying cluster structure. By combining rank estimation with our new low rank symmetric estimator and the graphical algorithm, we were able to recover most of the structure of the full system, with some of the weaker clusters being mixed up.

%\section*{Acknowledgments}
%We would like to acknowledge the assistance of volunteers in putting together this example manuscript and supplement.

\newpage

\bibliographystyle{siamplain}
%\bibliography{references}
%\bibliography{../References/bib_lib}
\bibliography{bib_lib}

\end{document}